\documentclass[12pt]{article}
\pdfoutput=1
\usepackage{epsf,amsfonts,amssymb,epsfig,amsmath,graphics,slashed}
\usepackage{hyperref,graphicx, subfig,hep}
\newcommand{\nn}{\notag \\}
\newcommand{\mub}{b}
\newcommand{\Bp}{\lambda}

\begin{document}

\makeatletter
\renewcommand{\theequation}{\thesection.\arabic{equation}}
\@addtoreset{equation}{section}
\makeatother

\baselineskip 18pt

\begin{titlepage}

\vfill

\begin{flushright}
Imperial/TP/2012/JG/06\\
\end{flushright}

\vfill

\begin{center}
   \baselineskip=16pt
   {\Large\bf  Semi-local quantum criticality in string/M-theory\\
     \vskip .2cm
}
  \vskip 1.5cm
      Aristomenis Donos, Jerome P. Gauntlett and Christiana Pantelidou\\
   \vskip .6cm
      \begin{small}
      \textit{Blackett Laboratory, 
        Imperial College\\ London, SW7 2AZ, U.K.}
        \end{small}\\*[.6cm]

\end{center}

\vfill

\begin{center}
\textbf{Abstract}
\end{center}

\begin{quote}
Semi-local quantum critical behaviour in $D-1$ spacetime dimensions can
be holographically described by metrics that are conformal to $AdS_2\times\mathbb{R}^{D-2}$, with the conformal
factor characterised by a parameter $\eta$. We analyse such ``$\eta$-geometries" 
in a top-down setting by focussing on the $U(1)^4$ truncation of $D=4$ $N=8$ gauged supergravity. 
The model has extremal black hole solutions carrying three non-zero electric or magnetic charges which approach $AdS_4$ in the UV
and an $\eta=1$ geometry in the IR. Adding a fourth charge provides a mechanism
to resolve the singularity of the $\eta$-geometry, replacing it with an $AdS_2\times\mathbb{R}^2$ factor in the IR, while maintaining
a large region where the $\eta$-geometry scaling is approximately valid. 
Some of the magnetically charged black hole solutions preserve supersymmetry while others just preserve it in the IR.
Finally, we show that $\eta$-geometries, with various values of $\eta$,
can be obtained from the dimensional reduction of geometries consisting of $AdS$ or Lifshitz geometries with
flat directions.
\end{quote}

\vfill

\end{titlepage}

\setcounter{equation}{0}


\section{Introduction}
A locally quantum critical fixed point exhibits a scaling of space and time of the form 
$t\to \lambda^z t$, ${\bf x}\to \lambda {\bf x}$ in the limit $z\to\infty$. 
Such scaling arises very naturally within the context of holography.
The simplest example is the standard electrically charge AdS-RN black brane solution of Einstein-Maxwell theory in $D$ spacetime dimensions.
At zero temperature, $T=0$, this solution interpolates between $AdS_D$ in the UV and an $AdS_2\times  \mathbb{R}^{D-2}$ fixed point solution
in the IR. From the dual perspective, this AdS-RN black hole solution describes a CFT in $D-1$ dimensions at finite charge density 
and the $T=0$ solution, providing it is stable, describes an emergent locally quantum critical fixed point in the far IR dual to the 
$AdS_2\times  \mathbb{R}^{D-2}$ solution. The correlators obtained from the $AdS_2\times  \mathbb{R}^{D-2}$ solution only scale with energy, but
since they still depend on the momentum we will follow \cite{Iqbal:2011in} and call this semi-local quantum criticality.
In the special case of $D=4$ the AdS-RN black hole geometry can also be supported by
magnetic fields or, more generally, by both electric and magnetic fields, with similar behaviour in the IR.

One interesting feature of the AdS-RN black holes is that they can have fermionic spectral functions with novel behaviour 
\cite{Lee:2008xf,Liu:2009dm,Cubrovic:2009ye,Faulkner:2009wj,Gauntlett:2011mf,Belliard:2011qq,Gauntlett:2011wm,DeWolfe:2011aa}. 
Another striking, and related,  feature is that they have finite entropy density at $T=0$. 
A natural interpretation is that this is indicating
that the system is becoming unstable at low temperatures and indeed, depending on the details of the gravitational system, there are a variety of possible superfluid and spatially modulated instabilities that can arise, both in bottom-up and top-down settings including
\cite{Gubser:2008px,Hartnoll:2008kx,Denef:2009tp,Gauntlett:2009dn,Gubser:2009gp,
Gubser:2008zu,Gubser:2008wv,Roberts:2008ns,Aprile:2010ge,Benini:2010qc,Nakamura:2009tf,Donos:2011bh,Donos:2011ff}. 
It is worth noting, however, that it has recently been shown that there is at least one top-down setting where the semi-local quantum critical ground state is known to be stable\footnote{At least in the strict $N\to \infty$ limit of the dual CFT.} via the preservation of supersymmetry \cite{Donos:2012sy}.

More recently it has been emphasised in \cite{Hartnoll:2012wm} that there is a broader framework to realise holographically the notion
of semi-local quantum criticality. The idea is to consider geometries that are conformally related to $AdS_2\times\mathbb{R}^{D-2}$. 
Specifically, we shall define the $\eta$-geometry by the line element
\begin{align}\label{etageom}
ds^2=\frac{1}{\rho^{2\eta/(D-2)}}\left(-\frac{dt^2}{\rho^2}+\ell^2 \frac{d\rho^2}{\rho^2}+dx_idx_i\right)\,,
\end{align}
where $i=1,\dots, D-2$, the number of spatial dimensions in the dual field theory, and $\eta$ and $\ell$ are constants.
Under the scaling $t\to\zeta t, x_i\to x_i, \rho\to \zeta\rho$ the line element scales as $ds\to \zeta^{-\eta/(D-2)}ds$.
The special case when $\eta=0$ is just $AdS_2\times\mathbb{R}^{D-2}$ with the conformal UV boundary located at $\rho\to 0$.
When $\eta> 0$ the geometry is singular when $\rho\to\infty$. Note that when $\eta=0$, $\ell$ is the radius of the $AdS_2$ factor and
that when
$\eta\ne 0$, $\ell$ can be set to one by scaling the coordinates.
One way to think about these geometries \cite{Hartnoll:2012wm} is as a limit of the ``hyperscaling violating" geometries considered in \cite{Charmousis:2010zz,Ogawa:2011bz,Huijse:2011ef}
labelled, in the notation of \cite{Huijse:2011ef}, by $\theta,z$ in the limit that $\theta\to-\infty$, $z\to\infty $ with $\eta\equiv  -z/\theta$ held fixed. Geometries with $\eta=1$ 
were also discussed earlier in \cite{Gubser:2009qt} and one of our aims will be to
generalise and extend the results of that pioneering paper (for related early work see \cite{Iizuka:2011hg,Gouteraux:2011ce} and more recent work \cite{DeWolfe:2012uv,Alishahiha:2012ad}).

An interesting property of the $\eta$-geometries is that the finite temperature generalisations have an entropy density that
depends on the temperature via $s\propto T^\eta$ and hence, when $\eta> 0$, they have $s\to 0$ as $T\to 0$ \cite{Gouteraux:2011ce,Hartnoll:2012wm}.
The case of $\eta=1$ is particularly interesting since the temperature dependence of $s$ is linear, corresponding to linear
specific heat \cite{Gubser:2009qt,Hartnoll:2012wm}. Another interesting feature when $\eta\ge 0$
is that there can be spectral weight that is not exponentially suppressed
at low-energies and finite momentum, as one expects for physics associated with Fermi surfaces \cite{Hartnoll:2012wm,Anantua:2012nj}.

In this paper we will explore several aspects of the $\eta$-geometries within top-down settings. 
The main focus will be on the $U(1)^4$ truncation of $D=4$ $N=8$ gauged supergravity \cite{Cvetic:1999xp}, studied earlier in this context in \cite{Gubser:2009qt}. 
Although this model is not quite a consistent truncation of $D=11$ supergravity on $S^7$, all of the solutions we consider can be uplifted
to obtain exact solutions in $D=11$. We show that for the class of
analytic black hole geometries with four electric charges found in \cite{Cvetic:1999xp}, or the magnetic analogues which we write down here,
we can obtain an $\eta=1$ geometry in the far
IR as $T\to 0$ after setting one of the charges to zero. In addition we show that introducing a small fourth charge resolves
the singularity with an $AdS_2\times\mathbb{R}^2$ geometry in the far IR, with an intermediate scaling region associated with an $\eta=1$ geometry.
This is reminiscent of the resolution of singularities in string theory by the addition of fluxes that have been considered in other contexts
\cite{Klebanov:2000hb} and also analogous to the resolution of the singularities of the Lifshitz geometries discussed in \cite{Harrison:2012vy}
(related work appears in \cite{Bhattacharya:2012zu,Kundu:2012jn,Bao:2012yt}).

We also consider the solutions after they are uplifted to $D=11$ on an $S^7$. For the uplifted electrically charged solutions we find the $\eta=1$ geometry region uplifts to a $D=11$ solution with an $AdS_3$ factor, 
generalising what was seen in \cite{Gubser:2009qt}. The presence of the $AdS_3$ factor provides an understanding of the linear specific heat 
\cite{Gubser:2009qt}. However, this is not the full story since, 
by contrast, we show that there is no such $AdS_3$ factor in the uplifted
magnetically charged solutions.

We will not carry out a detailed stability analysis of these analytic black hole solutions here. However, based on 
\cite{Gubser:2000mm} and on the
analysis of the stability properties of the $AdS_2\times\mathbb{R}^2$ geometries presented in \cite{Donos:2011pn},
we expect that many of the analytic black hole solutions, carrying either electric or magnetic charges, are unstable. Such instabilities are certainly interesting since they are associated with new branches of black hole solutions appearing
at finite temperature, corresponding to new phases. However, such instabilities also mean that many of the $\eta=1$ geometries will probably not correspond to the true ground states at zero temperature.

On the other hand there is a particularly interesting subclass of the analytic black hole solutions where the
instabilities are ameliorated by the
presence of an emergent supersymmetry in the IR. 
This subclass has four non-vanishing magnetic charges and while in the extremal $T=0$ limit they
are not supersymmetric solutions they nevertheless approach supersymmetric $AdS_2\times\mathbb{R}^2$ geometries in the IR
of the type constructed in \cite{Donos:2011pn,Almheiri:2011cb} (building on \cite{Almuhairi:2011ws}). This emergent supersymmetry
implies that the near horizon region is free from instabilities and suggests that the full solutions themselves may also
be stable. If this is the case, these solutions would provide the first examples of stable, non-supersymmetric black brane solutions with finite
entropy at zero-temperature. Moreover, these solutions can exist with an approximate intermediary $\eta=1$ geometry scaling region which dominates the IR when one of the charges is set to zero.

Another result of this paper is that the same $D=4$ $U(1)^4$ theory also admits a new
class of solutions carrying purely magnetic charges with analogous properties to those described in the previous paragraph,
but preserving supersymmetry everywhere. In particular, we numerically construct supersymmetric solutions
interpolating between $AdS_4$ in the UV and $\eta$-geometries with $\eta=1$ in the IR.
Adding small amounts of a fourth charge again provides a natural singularity resolution mechanism with 
an intermediate $\eta=1$ geometry scaling region and a supersymmetric $AdS_2\times\mathbb{R}^2$ solution in the far IR. A duality transformation maps these supersymmetric magnetic solutions to a new class of non-supersymmetric electric solutions.

Finally, in a quite different direction, we conclude the paper by briefly showing that a simple way to construct $\eta$-geometries is from the dimensional reduction of the product of $AdS$ or Lifshitz geometries with some flat directions.
Similar observations were made earlier (independently) in the context of specific classes of models in
\cite{Gouteraux:2011ce}.

The plan of the rest of the paper is as follows. In section \ref{the model} we introduce the $D=4$ $U(1)^4$ model
that we mostly consider, and also recall the magnetic and electrically charged 
$AdS_2\times \mathbb{R}^2$ solutions of \cite{Donos:2011pn}.
In section \ref{sec:an_sol} we analyse the analytic class of asymptotically AdS$_4$ black brane solutions carrying electric charges found 
in \cite{Cvetic:1999xp}.  
We discuss the analytic magnetically charged black holes in section \ref{magsec1} 
and the numerically constructed supersymmetric magnetic solutions in section \ref{magsec2}. 
We conclude in section \ref{dimredsec} by
obtaining the $\eta$-geometries via dimensional reduction.

\section{The $D=4$ gauged supergravity theory}\label{the model}
We consider the $U(1)^4$ truncation of $N=8$ $D=4$ gauged supergravity given in \cite{Cvetic:1999xp}
that keeps three neutral scalar fields $\phi_a$ and four gauge fields $A^i$. Solutions of this theory will be the major focus of the paper. 
The Lagrangian is given by
\begin{align}\label{eq:4dLag}
\mathcal{L}&=\frac{1}{2}R-\frac{1}{4}\,\sum_{a=1}^{3}\left(\partial\phi_{a} \right)^{2}-\sum_{i=1}^{4} X_{i}^{-2}\,\left(F^{i}\right)_{\mu\nu}\left(F^{i}\right)^{\mu\nu}-V\left(X_{i} \right)\,,
\end{align}
where
\begin{align}
&X_{1}=e^{\frac{1}{2}\left(-\phi_{1}-\phi_{2}-\phi_{3} \right)},\quad X_{2}=e^{\frac{1}{2}\left(-\phi_{1}+\phi_{2}+\phi_{3} \right)},\quad X_{3}=e^{\frac{1}{2}\left(\phi_{1}-\phi_{2}+\phi_{3} \right)},\quad X_{4}=e^{\frac{1}{2}\left(\phi_{1}+\phi_{2}-\phi_{3} \right)}
\end{align}
and the potential is given by
\begin{align}
&V\left(X_{i} \right)=-\frac{1}{2}\,\sum_{i\neq j}\,X_{i}X_{j}=-2\left(\cosh\phi_1+\cosh\phi_2+\cosh\phi_3\right)\,.
\end{align}
Any solution of this theory that satisfies $F^{i}\wedge F^{j}=0$ can be 
uplifted\footnote{To do this we should set $g^2=1/2$ in eq. (3.8) of \cite{Cvetic:1999xp}
and identify $(F^{i})^{there}=2\sqrt{2}(F^{i})^{here}$. It is also worth noting that we are using the same conventions as in
\cite{Duff:1999gh} setting $g=1$ there.}
to $D=11$ on an $S^7$ using the formulae in \cite{Cvetic:1999xp};  all of the solutions that we consider satisfy this condition.

Note that the equations of motion for this model exhibit the electric-magnetic duality transformation given by
\begin{align}\label{dual}
F^i\to X_i^{-2}*F^i,\qquad \phi_a\to -\phi_a\,,
\end{align}
with the metric unchanged.

In the following we will sometimes utilise the fact that the equations of motion for \eqref{eq:4dLag} can be consistently truncated
to theories involving a smaller numbers of fields. For example it is consistent to further truncate by setting 
\begin{align}\label{firsttrunc}
&\phi_2=-\phi_3,\qquad\textit{i.e.}\qquad X_1=X_2\,,\nn
&F^1=F^2\,,
\end{align}
to obtain a theory with two scalar fields and three vector fields. This should be a sector of an $SU(2)\times 
U(1)\times U(1)$ invariant subsector of $SO(8)$
gauged supergravity.
On the other hand we can set
\begin{align}\label{second}
&\phi_1=\phi_2=-\phi_3,\qquad\textit{i.e.}\qquad X_1=X_2=X_3\,,\nn
&F^1=F^2=F^3\,,
\end{align}
to obtain a theory involving one scalar field and two gauge-fields. In fact this is a 
sector of the $SU(3)$ invariant subsector of $SO(8)$
gauged supergravity \cite{Warner:1983vz}\cite{Bobev:2010ib} and the corresponding uplifted solutions will have $SU(3)\times U(1)^2$ symmetry.

This theory has an $AdS_4$ vacuum, with with $\phi_a=0$ radius squared 1/2, 
which can be uplifted to $D=11$ to give the
$AdS_4\times S^7$ solution. In this $AdS_4$ vacuum the three neutral scalars have $m^2=-4$ and
hence can be quantised as $\Delta=1$ or $\Delta=2$ operators. For the $AdS_4\times S^7$ solution
to be consistent with supersymmetry, they should be quantised so that $\Delta=1$ (for more discussion see e.g. \cite{Donos:2011ut}). There may be sub-truncations and/or other uplifts where it is appropriate to quantise as a $\Delta=2$ operator, but we will continue assuming $\Delta=1$.

\subsection{Analytic $AdS_2\times \mathbb{R}^2$ solutions}
We briefly review the $AdS_2\times \mathbb{R}^2$ solutions supported by magnetic or electric charges that were studied in \cite{Donos:2011pn,Almheiri:2011cb}
as they will appear in the subsequent analysis.

\subsubsection{Magnetic $AdS_2\times \mathbb{R}^2$ solutions}\label{magads2sols}
The solutions supported by magnetic flux are given by
\begin{align}\label{eq:AdS2ansatz}
ds^{2}&=L^{2}\,ds^{2}\left(AdS_{2}\right)+dx_{1}^{2}+dx_{2}^{2}\,,\notag\\
F^{i}&=\tfrac{1}{2}q_i\,dx_{1}\wedge dx_{2}\,,\nn
\phi_1&=f_1,\quad \phi_2=f_2,\quad\phi_3=f_3\,,
\end{align}
where $q_i$, $f_a$ are constants and $L$ is the $AdS_2$ radius.
Defining the on-shell quantities
\begin{align}\label{exs}
\bar X_{1}&=e^{\frac{1}{2}\left(-f_{1}-f_{2}-f_{3} \right)},\quad \bar X_{2}=e^{\frac{1}{2}\left(-f_{1}+f_{2}+f_{3} \right)},\quad \bar X_{3}=e^{\frac{1}{2}\left(f_{1}-f_{2}+f_{3} \right)},\quad \bar X_{4}=e^{\frac{1}{2}\left(f_{1}+f_{2}-f_{3} \right)}\,,
\end{align}
there is a three parameter family of solutions specified by arbitrary values of $(f_1,f_2,f_3)$ with 
\begin{align}\label{eq:ads2sol}
q_i^{2}=\frac{\bar{X}_{i}^{2}}{2}\,\sum_{j\neq k\neq i} \bar{X}_{j}\bar{X}_{k},\qquad
L^{-2}=-2\, V\left(\bar{X}_{i} \right)\,.
\end{align}
Note that the $q_i$ can be chosen to have either sign.
In order for these solutions to preserve supersymmetry it is necessary that one of the following conditions is satisfied:
\begin{align}\label{omeq}
q_1+q_2+q_3+q_4=0, \qquad q_1+q_2-q_3-q_4=0,\nn q_1-q_2+q_3-q_4=0,\qquad  q_1-q_2-q_3+q_4=0.
\end{align}
Furthermore, it was shown in \cite{Donos:2011pn} that this implies that the supersymmetry locus is given by the conditions
\begin{align}\label{susloc}
2\sum \bar X_i^2=\left(\sum_i\bar X_i\right)^2\,.
\end{align}
In fact, conversely, this condition, along with \eqref{eq:ads2sol} implies the preservation of supersymmetry.
Indeed, \eqref{susloc} and \eqref{eq:ads2sol} imply that $2q_i=\pm \bar X_i(-2\bar X_i+\sum_j\bar X_j)$
which, along with demanding one of the conditions in \eqref{omeq}, is sufficient for preservation of supersymmetry.

\subsubsection{Electric $AdS_2\times \mathbb{R}^2$ solutions}\label{elads2sols}
The solutions supported by electric flux can be obtained from the duality transformation
\eqref{dual}. Explicitly they are given by
\begin{align}\label{eq:AdS2ansatze}
ds^{2}&=L^{2}\,ds^{2}\left(AdS_{2}\right)+dx_{1}^{2}+dx_{2}^{2}\,,\notag\\
F^{i}&=\tfrac{1}{2}q_iL^2Vol(AdS_2)\,,\nn
\phi_1&=f_1,\quad \phi_2=f_2,\quad\phi_3=f_3\,,
\end{align}
where 
\begin{align}\label{elads2exp}
q_i=\bar{X}_{i}^{3}\sum_{j\neq i} \bar{X}_{j},\qquad
L^{-2}=-2\, V\left(\bar{X}_{i} \right)\,,
\end{align}
and the $\bar X_{i}$ are the on-shell quantities defined in \eqref{exs}. These solutions do not
preserve supersymmetry.

\section{Electric black hole solutions and $\eta$-geometries}\label{sec:an_sol}
We begin with the analytic class of asymptotically AdS$_4$ black brane solutions carrying four electric charges \cite{Cvetic:1999xp}
\begin{align}\label{eq:an_sol}
ds^{2}&=-f\,\Pi^{-1}\,dt^{2}+f^{-1}\,\Pi\,dr^{2}+r^{2}\,\Pi\,\left( dx_{1}^{2}+dx_{2}^{2}\right)\nn
A^{i}&=\frac{\varepsilon_i}{2}\,\left(\mu_{i}+\frac{1}{\sqrt{2\,Q_{i}}}\,\left(1-H_{i}^{-1}\right)\right)\,dt,
\qquad X_{i}=H_{i}^{-1}\,\Pi^{1/2},
\end{align}
where
\begin{align}\label{fhp}
f=-\frac{\mub}{r}+2r^{2}\Pi^{2}, \qquad H_{i}=1+\frac{\mub Q_{i}}{r},
\qquad \Pi=\left(H_{1}H_{2}H_{3}H_{4} \right)^{1/2}\,.
\end{align}
We have $Q_i\ge 0$, $b\ge 0$ and $\varepsilon_i=\pm 1$.
As $r\to\infty$ the solutions approach $AdS_4$. The black hole event horizon is located at $r=r_h\ge 0$ where
$r_{h}$ is the largest root of the equation
\begin{align}\label{bhcond}
(\Pi^{-1}f)(r_h)=0\,.
\end{align}
The chemical potentials, $\mu_{i}$, for the four $U(1)$'s are given by
\begin{align}\label{eq:chem_pot}
\mu_{i}=\frac{1}{\sqrt{2\,Q_{i}}}\,\left(H_{i}^{-1}\left(r_{h}\right)-1\right)\,,
\end{align}
ensuring regularity of the gauge-potentials at the horizon.
Note that $\mub$ is the ``emblackening" parameter (we will not denote it as $\mu$, as is often done, to avoid confusion with the chemical potentials in the dual CFT).

In order to analyse the asymptotic UV behaviour of the scalar fields, it is useful
to introduce a new radial coordinate $\rho^2=r^2\Pi$. We then find as $\rho\to\infty$ 
\begin{align}
\phi_1&= \frac{b(Q_1+Q_2-Q_3-Q_4)}{2\rho}-\frac{b^2(Q_1-Q_2+Q_3-Q_4)(Q_1-Q_2-Q_3+Q_4)}{8\rho^2}+\dots\nn
\phi_2&= \frac{b(Q_1-Q_2+Q_3-Q_4)}{2\rho}-\frac{b^2(Q_1+Q_2-Q_3-Q_4)(Q_1-Q_2-Q_3+Q_4)}{8\rho^2}+\dots\nn
\phi_3&= \frac{b(Q_1-Q_2-Q_3+Q_4)}{2\rho}-\frac{b^2(Q_1+Q_2-Q_3-Q_4)(Q_1-Q_2+Q_3-Q_4)}{8\rho^2}+\dots\
\end{align}
Thus for the $\Delta=1$ quantisation relevant for maximal supersymmetry, 
we see that, generically, there are non-zero deformations, corresponding to the $1/\rho^2$ pieces and 
non-zero expectation values, corresponding to the $1/\rho$ pieces (assuming that there is no mixing).
It is also worth noting 
that if the $Q_i$ are chosen so that one of the deformation parameters vanishes, then both of the other
two expectation values do as well.

Notice that these analytic black hole solutions depend on 5 independent parameters: four $\mu_i$ and $b$ (the
$Q_i$ are fixed by regularity at the black hole event horizon).
If we stay within a static, spatially homogeneous and isotropic context, and with electric charges only,
the most general solutions should depend on 8 parameters (and there could be discrete families of solutions). 
These can be viewed as the temperature,
four chemical potentials $\mu_i$ and three deformations for the $\Delta=1$ operator (or the $\Delta=2$
operator in the other quantisation).
In section \ref{magsec2} we will numerically construct some new solutions outside of the analytic family.

\subsection{Four $Q_i\ne0$: $AdS_2\times\mathbb{R}^2$ in the IR at $T=0$}\label{elads2}

Let us first consider the generic case when all four of the $Q_i$ are non-zero. We will show that in
the extremal, $T=0$, limit the black hole solutions all approach smooth domain wall solutions interpolating between
$AdS_4$ in the UV and $AdS_2\times\mathbb{R}^2$ in the IR. In particular, all of these black hole solutions have finite entropy at $T=0$. Furthermore, we will see that the entire
moduli space of electrically charged $AdS_2\times\mathbb{R}^2$ solutions given in section \ref{elads2sols}
can be obtained.

To begin with we rescale the radial coordinate via $r\rightarrow b\,\rho$. For an extremal black hole event horizon, in addition to \eqref{bhcond} we have $ (\Pi^{-1}f)'(r_h)=0$. These conditions imply the relations
\begin{align}\label{eq:el_const_q}
b^2&=\frac{\rho_{h}}{2\,\left(Q_{1}+\rho_{h}\right)\left(Q_{2}+\rho_{h}\right)\left(Q_{3}+\rho_{h}\right)\left(Q_{4}+\rho_{h}\right)}\,,\nn
Q_{4}&=\rho_{h}^{2}\frac{ Q_1Q_2+Q_1 Q_3+Q_2Q_3 + 2\rho_h(Q_1+Q_2+Q_3)+3\rho_h^2}{Q_{1}Q_{2}Q_{3}-\rho_h^2\left(Q_{1}+Q_{2}+Q_{3} \right)-2\rho_{h}^{3}}\,,
\end{align}
The second equation fixes $Q_{4}$ in terms of $Q_1,Q_2,Q_3$ and also the location of the extremal horizon at $\rho=\rho_{h}$. It is convenient now to rescale the charges
\begin{align}
Q_i=\rho_h \bar q_i^2\,,
\end{align}
and upon evaluating the scalars on the horizon we find
\begin{align}\label{eq:phi_hor_el}
e^{-2\phi_{1}}&=\frac{\left(1+\bar q_{3}^{2}\right)^{2}}{\bar q_{1}^{2}\bar q_{2}^{2}\bar q_{3}^{2}-2-\bar q_{1}^{2}-\bar q_{2}^{2}-\bar q_{3}^{2}}\,,\nn
e^{-2\phi_{2}}&=\frac{\left(1+\bar q_{2}^{2}\right)^{2}}{\bar q_{1}^{2}\bar q_{2}^{2}\bar q_{3}^{2}-2-\bar q_{1}^{2}-\bar q_{2}^{2}-\bar q_{3}^{2}}\,,\nn
e^{2\phi_{3}}&=\frac{\left(1+\bar q_{1}^{2}\right)^{2}}{\bar q_{1}^{2}\bar q_{2}^{2}\bar q_{3}^{2}-2-\bar q_{1}^{2}-\bar q_{2}^{2}-\bar q_{3}^{2}}\,.
\end{align}
Notice that the condition for the positivity of $Q_{4}$ (i.e. the reality of $\bar q_4$) 
is the same as that for the reality of $\phi_{a}$ in \eqref{eq:phi_hor_el}. It is now easy to invert equation \eqref{eq:phi_hor_el} and express the constants $\bar q_1$, $\bar q_2$, $\bar q_3$ 
in terms of the scalars $\phi_{a}$ and we find
\begin{align}\label{firstcondsq}
\bar q_{1}^{2}=\frac{1}{X_{1}}\left(X_2+X_3+X_4\right)\,,\nn
\bar q_{2}^{2}=\frac{1}{X_{2}}\left(X_1+X_3+X_4\right)\,,\nn
\bar q_{3}^{2}=\frac{1}{X_{3}}\left(X_1+X_2+X_4\right)\,.
\end{align}
Analysing the behaviour of the metric we obtain
\begin{align}\label{neesol}
ds^2=-\frac{b^2(\rho-\rho_h)^2}{L^2}dt^2
+\frac{L^2}{(\rho-\rho_h)^2}{d\rho^2}+\frac{b\rho_h^{1/2}}{\sqrt 2}(dx_1^2+ dx_2^2)\,,
\end{align}
with $L^{-2}$ as in \eqref{elads2exp}. Analysing the flux in the near horizon limit we also
obtain the same expression in \eqref{elads2exp} after identifying 
\begin{align}
q_i=X_1^2 \bar q_i\,.
\end{align}
Finally, we can check that the conditions \eqref{firstcondsq} are now precisely as in
\eqref{elads2exp}. In other words we have shown that we can obtain the full moduli space of
electric $AdS_{2}\times \mathbb{R}^{2}$ solutions of \cite{Donos:2011pn}.

\subsection{Three $Q_i\ne 0$: Geometries with $\eta=1$ in the IR at $T=0$}
Next we consider the special case that one of the four charges is zero. For definiteness we choose $Q_4=0$. 
As we can see from \eqref{eq:el_const_q}, the extremal $T=0$ limit is achieved when 
$\mub=\frac{1}{\sqrt{2Q_{1}Q_{2}Q_{3}}}$ with $r_h\to 0$.
In the near horizon limit, as $r\to 0$, the geometry now approaches
\begin{align}\label{etone}
ds^{2}&\approx -U_0\,r^{3/2}\,dt^{2}+U_0^{-1}\,\frac{dr^{2}}{r^{3/2}}+\frac{r^{1/2}}{\left(8 Q_{1}Q_{2}Q_{3} \right)^{1/4}}\,\left(dx_{1}^{2}+dx_{2}^{2} \right)\,,\nn
U_0&=\frac{Q_{1}Q_{2}+Q_{1}Q_{3}+Q_{2}Q_{3}}{\left(\frac{1}{2}Q_{1}Q_{2}Q_{3} \right)^{3/4}}\,,
\end{align}
while the scalars approach
\begin{align}\label{scalapp}
\phi_{1}&\approx\frac{1}{4}\,\ln\left(\frac{Q_{1}Q_{2}}{2Q^{3}_{3}} \right)-\frac{1}{2}\,\ln r\,,\nn
\phi_{2}&\approx\frac{1}{4}\,\ln\left(\frac{Q_{1}Q_{3}}{2Q^{3}_{2}} \right)-\frac{1}{2}\,\ln r\,,\nn
\phi_{3}&\approx-\frac{1}{4}\,\ln\left(\frac{Q_{2}Q_{3}}{2Q^{3}_{1}} \right)+\frac{1}{2}\,\ln r\,.
\end{align}
We also find that the three non-trivial gauge-fields can be written
\begin{align}\label{gfet}
A_{i}&\approx-\frac{\varepsilon_{i}}{2}\,\frac{Q_1Q_2Q_3}{Q_{i}^{3/2}}r\,dt\,,\qquad i=1,2,3\,,
\end{align}
where we have used the fact that when $Q_4=0$ the chemical potentials defined in \eqref{eq:chem_pot} are simply $\mu_i=-1/(2Q_i)^{1/2}$.
After the coordinate change
\begin{align}
 t\rightarrow \frac{8}{U_0^{2}}\,t,\qquad r\rightarrow \frac{U_0^{2}}{16}\,\rho^{-2},\qquad x_{i}\rightarrow \frac{2\,\left(8 Q_{1}Q_{2}Q_{3} \right)^{1/8}}{U^{1/2}}\,x_{i}\,,
\end{align}
we see that we get a semi-local quantum critical metric with $\eta=1$:\begin{align}
ds^{2}\approx \frac{1}{\rho}\,\left(-\frac{dt^{2}}{\rho^{2}}+\frac{d\rho^{2}}{\rho^{2}}+dx_{1}^{2}+dx_{2}^{2} \right)\,.
\end{align}
It is worth emphasising that the $\eta=1$ geometry is not an exact solution of the equations of motion.

We expect that the $Q_4=0$ solutions at finite temperature have an entropy that behaves as $s\to T$, for low temperatures
\cite{Hartnoll:2012wm}, corresponding to linear specific heat. We can see this behaviour as follows. For small temperatures the horizon will be located at $r=\delta r_{h}$. Since we require that 
the chemical potentials \eqref{eq:chem_pot} are fixed we deduce that
\begin{equation}
\delta Q_{i}=-2\sqrt{2Q_{1}Q_{2}Q_{3}}\,\delta r_{h}\,,\qquad i=1,2,3\,.
\end{equation}
On the other hand, using this and the definition of the location of the black hole event horizon \eqref{bhcond}, we conclude that
we should vary $\mub$ according to
\begin{equation}
\delta\mub=\frac{Q_{1}Q_{2}+Q_{1}Q_{3}+Q_{2}Q_{3}}{2Q_{1}Q_{2}Q_{3}}\,\delta r_{h}\,.
\end{equation}
Recalling the definitions of the temperature and entropy density (with $16\pi G=2$)
\begin{align}
T=\left.\frac{\left(f\,\Pi^{-1}\right)^{\prime}}{4\pi}\right|_{r=r_{h}},\qquad
s=2\pi\,\left. r^{2}\Pi\right|_{r= r_{h}}\,,
\end{align}
we deduce, at leading order in the variations, that
\begin{equation}
\delta T=\frac{Q_{1}Q_{2}+Q_{1}Q_{3}+Q_{2}Q_{3}}{2^{5/4}\pi\left(Q_{1}Q_{2}Q_{3} \right)^{3/4}}\,\sqrt{\delta r_{h}}\,,\qquad
\delta s=\frac{2^{1/4}\pi}{\left(Q_{1}Q_{2}Q_{3} \right)^{1/4}}\,\sqrt{\delta r_{h}}\,,
\end{equation}
and hence
\begin{equation}
\delta s=\frac{2\sqrt{2}\pi^{2}\sqrt{Q_{1}Q_{2}Q_{3}}}{Q_{1}Q_{2}+Q_{1}Q_{3}+Q_{2}Q_{3}}\,\delta T\,,
\end{equation}
as claimed.

\subsection{$AdS_3$ in the uplift}
If we uplift this entire class of geometries with $Q_4=0$, we find that the $\eta=1$ geometry appearing in the IR at $T=0$ uplifts to a
locally $AdS_3$ region. This generalises the result of \cite{Gubser:2009qt} which considered the special case $Q_1=Q_2=Q_3$. It was
also pointed out in \cite{Gubser:2009qt} that the $AdS_3$ factor provides a natural interpretation of the behaviour $s\propto T$ that we saw in the last subsection.

Specifically, if we uplift the $\eta=1$ limiting IR geometry \eqref{etone}, \eqref{scalapp}, \eqref{gfet} to eleven dimensions using \cite{Cvetic:1999xp} (see footnote 1) we obtain, as $r \to 0$,
\begin{align}\label{elul}
&ds^{2}_{11}\approx \frac{1}{\left(8Q_1Q_2Q_3 \right)^{1/12}}\,\mu_{4}^{4/3}\,\left(-U_{0}r\,dt^{2}+U_{0}^{-1}\,\frac{dr^{2}}{r^{2}}+2\,\left(8Q_1Q_2Q_3 \right)^{1/4}r\,d\phi_{4}^{2} \right)\nn
&+\frac{\mu_{4}^{4/3}}{\left(8Q_1Q_2Q_3\right)^{1/3}}\,\left(dx_{1}^{2}+dx_{2}^{2} \right)+\frac{2}{(Q_1Q_2Q_3)^{1/3}}\,\mu_{4}^{-2/3}\,\sum_{i=1}^{3}\,Q_{i}\,\left( d\mu_{i}^{2}+\mu_{i}^{2}\,\left(d\phi_{i}+2 A_{i}\,dt \right)^{2}\right),
\end{align}
where $A_i$ are given in \eqref{gfet}. It is interesting to point out that the chemical potentials for the gauge-fields that we have used, which arose from
regularity at the event horizon at finite temperature, imply that the metrics are free of closed time-like curves in $D=11$, in contrast to
the gauge used in \cite{Gubser:2009qt}.

\subsection{Charge as a resolution mechanism and intermediate scaling}\label{sec:an_interm}
We have shown that the class of solutions with $Q_4=0$ at $T=0$ all approach an $\eta=1$ geometry in the IR and hence are singular.
On the other hand we showed in section \ref{elads2} that when all four charges are non-zero the solutions approach $AdS_2\times\mathbb{R}^2$ 
in the IR. It is thus clear that that adding a small fourth charge, $Q_4\ne 0$,
will resolve the $\eta$-geometry singularity. In addition, for small $Q_4$ we expect to
obtain an intermediate scaling regime where the geometry is essentially
the $\eta$-geometry and then very far in the IR, it approaches the $AdS_2\times \mathbb{R}^2$ solution. 
This is analogous to the singularity resolution of Lifshitz geometries discussed in \cite{Harrison:2012vy}.

To illustrate this point in more detail, for simplicity we now focus on the sub-class of extremal solutions
with $Q_{1}=Q_{2}=Q_{3}\equiv Q/\mub$ and $Q_{4}\equiv q/\mub$, with $b=2^{-1/2}Q_1^{-3/2}$. 
The fourth charge, $Q_4$, will be much smaller than the other three if $q<<Q$.
It is convenient to parametrise the family of solutions in terms of the location of the extremal horizon, $r=r_{h}$.
Doing so we obtain the relation
\begin{equation}
q=\frac{3\,r_{h}^{2}}{Q-2\,r_{h}}\,,
\end{equation}
while the metric reads
\begin{align}
ds^{2}&=-U\,dt^{2}+U^{-1}\,dr^{2}+W\,\left(dx_{1}^{2}+dx_{2}^{2} \right)\,,\nn
W&=\left(Q+r \right)^{3/2}\,\left(\frac{3\,r_{h}^{2}}{Q-2r_{h}}+r\right)^{1/2}\,,\nn
U&=2\,\left(r-r_{h}\right)^{2}\,\frac{3Q^{3}+3Q^{2}r+Qr\left(r-4\,r_{h} \right)-r\,r_{h}\,\left(2r+r_{h}\right)}{\left(Q-2r_{h} \right)\left(Q+r\right)^{3/2}\left(\frac{3\,r_{h}^{2}}{Q-2r_{h}}+r\right)^{1/2}}\,.
\end{align}
We see that when $r_{h}>0$ we have an $AdS_{2}\times \mathbb{R}^{2}$ geometry in the IR. On the other hand
when $r_{h}=0$, we have $Q_4=0$ and we are back in the situation that we described in section \ref{sec:an_sol} for the special
case of three equal non-vanishing charges. In particular we obtain the $\eta=1$ geometry \eqref{etone} as $r\to 0$.

In order to illustrate the intermediate scaling region it is illuminating to define the functions
\begin{align}
p_{1}=\left(r-r_{h} \right)\,\frac{U^{\prime}}{U}\,,\qquad
p_{2}=\left(r-r_{h} \right)\,\frac{W^{\prime}}{W}\,,
\end{align}
and explicitly we have 
\begin{align}
p_{1}=&2+\frac{3 (Q+r_{h})}{2 (Q+r_{h}+y)}+\frac{r_{h} (Q+r_{h})}{2 r_{h} (r_{h}-2 y)+2 Q (r_{h}+y)}\nn
&\qquad\qquad-\frac{(Q+r_{h}) \left(6 Q^2+3 Q y-r_{h} (6 r_{h}+5 y)\right)}{3 (Q-r_{h}) (Q+r_{h})^2+(3 Q-5 r_{h}) (Q+r_{h}) y+(Q-2 r_{h}) y^2}\,,\nn
p_{2}=&\frac{y \left((Q+r_{h})^2+4 (Q-2\,r_{h}) y\right)}{2 (Q+r_{h}+y) (r_{h} (r_{h}-2\,y)+Q (r_{h}+y))}\,,
\end{align}
with $y=r-r_{h}$.
We now focus on three different scaling regions obtaining
\begin{align}
p_{1}&\approx\begin{cases}
2,& y\rightarrow 0\\
3/2, & r_{h} << y<< Q\\
2,& y>>Q>>r_{h}
\end{cases}\nn
p_{2}&\approx\begin{cases}
0,& y\rightarrow 0\\
1/2, & r_{h} << y<< Q\\
2, & y>>Q>>r_{h}\,.
\end{cases}
\end{align}
For $r_h\ne 0$, as $y\to 0$ we see the scaling behaviour of the
$AdS_2\times\mathbb{R}^2$ geometry. Similarly for very large $y$ we see the scaling associated with
the asymptotic $AdS_4$ geometry.  Finally, when an intermediate region with $r_{h}<< y << Q$ exists, the metric has the scaling behaviour of an $\eta=1$ geometry
(see \eqref{etone}). As expected such a region exists 
for $q<<Q$.

Similar observations also hold for the three scalar fields. To see this we first recall from \eqref{second}
that the sub-class of solutions with three equal charges that we are focussing on are actually solutions of a consistent truncation of the equations of motion of
\eqref{eq:4dLag} to a theory with a single scalar field: $\phi=\phi_{1}=\phi_{2}=-\phi_{3}$ and two vector fields. 
We therefore examine the quantity
\begin{align}
p_{3}=\left(r-r_{h}\right)\,\phi^{\prime}=-\frac{(Q-3 r_{h}) (Q+r_{h}) y}{2 (Q+r_{h}+y) (r_{h} (r_{h}-2 y)+Q (r_{h}+y))}\,,
\end{align}
which has the behaviour
\begin{equation}
p_{3}\approx \begin{cases}
0,& y\rightarrow 0\\
-1/2, & r_{h} << y<< Q\\
0, & y>>Q>>r_{h}\,.
\end{cases}
\end{equation}
In the intermediate scaling region we again see the behaviour expected for an $\eta=1$ geometry (see \eqref{etone},\eqref{scalapp}).

\section{Analytic magnetically charged black holes}\label{magsec1}
The magnetic version of the analytic black hole solutions is easily obtained from the analytic electric solutions
\eqref{eq:an_sol} using the duality transformation \eqref{dual}. Explicitly
we have
\begin{align}\label{eq:an_soltwo}
ds^{2}&=-f\,\Pi^{-1}\,dt^{2}+f^{-1}\,\Pi\,dr^{2}+r^{2}\,\Pi\,\left( dx_{1}^{2}+dx_{2}^{2}\right)\,,\nn
F_{i}&=-\varepsilon_{i}\,\frac{\sqrt{Q_{i}}\mub}{2\sqrt{2}}\,dx_{1}\wedge dx_{2}\,,
\qquad X_{i}=H_{i}\,\Pi^{-1/2},
\end{align}
where $f,H_i$ and $\Pi$ are the same as \eqref{fhp}. Since the metric is unchanged, many of the properties
we saw in the previous section for the electric solutions follow straightforwardly.

In particular, when three magnetic charges are non-zero we obtain $\eta$-geometries with $\eta=1$ at $T=0$ in the far IR.
Furthermore, when we switch on a small fourth magnetic charge we obtain solutions at $T=0$ that have an
intermediate scaling region associated with an $\eta=1$ geometry and then in the far IR
approach a magnetically charged $AdS_2\times\mathbb{R}^2$ solution of section \ref{magads2sols}.

Using the supersymmetry transformations given in section 5, we can deduce that, as for
the electric solutions discussed in the last section, these analytic magnetic solutions do not preserve any
supersymmetry. However, they can exhibit an interesting emergent supersymmetry at $T=0$ in the far IR as we now explain.

When all four magnetic charges are non-zero the analysis of the near horizon limit in the extremal $T=0$ case is almost identical to the electric case that we considered in section \ref{elads2}. Rescaling $r\rightarrow b\,\rho$ the extremal
$T=0$ limit of the solutions \eqref{eq:an_soltwo} again lead to the
conditions
\eqref{eq:el_const_q}. We next scale the magnetic charges via
\begin{align}
Q_i=\rho_h q_i^2\,,
\end{align}
where now $q_i$ are the magnetic charges appearing in the magnetic $AdS_2\times\mathbb{R}^2$ solutions given
in \eqref{eq:AdS2ansatz}.
Evaluating the scalars on the horizon we have
\begin{align}\label{eq:phi_hor}
e^{2\phi_{1}}&=\frac{\left(1+q_{3}^{2}\right)^{2}}{q_{1}^{2}q_{2}^{2}q_{3}^{2}-2-q_{1}^{2}-q_{2}^{2}-q_{3}^{2}}\,,\nn
e^{2\phi_{2}}&=\frac{\left(1+q_{2}^{2}\right)^{2}}{q_{1}^{2}q_{2}^{2}q_{3}^{2}-2-q_{1}^{2}-q_{2}^{2}-q_{3}^{2}}\,,\nn
e^{-2\phi_{3}}&=\frac{\left(1+q_{1}^{2}\right)^{2}}{q_{1}^{2}q_{2}^{2}q_{3}^{2}-2-q_{1}^{2}-q_{2}^{2}-q_{3}^{2}}\,,
\end{align}
and hence
\begin{align}
q_{1}^{2}=\frac{X_{1}}{X_{4}}+\frac{X_{1}}{X_{2}}+\frac{X_{1}}{X_{3}}\,,\nn
q_{2}^{2}=\frac{X_{2}}{X_{1}}+\frac{X_{2}}{X_{4}}+\frac{X_{2}}{X_{3}}\,,\nn
q_{3}^{2}=\frac{X_{3}}{X_{2}}+\frac{X_{3}}{X_{1}}+\frac{X_{3}}{X_{4}}\,.
\end{align}
We observe that these are precisely the same conditions appearing in \eqref{eq:ads2sol}.
Since the metric is as in \eqref{neesol}, we conclude that we can obtain all of the magnetic 
$AdS_{2}\times \mathbb{R}^{2}$ solutions of \cite{Donos:2011pn} in the IR.

In particular, the sub-locus of the magnetic $AdS_{2}\times \mathbb{R}^{2}$ solutions of \cite{Donos:2011pn} that preserve supersymmetry, i.e. satisfying
\eqref{eq:ads2sol} and \eqref{susloc},
can
be obtained as near horizon limits of non-supersymmetric extremal black hole solutions. This emergent
supersymmetry is interesting. One consequence is that the black hole solutions must be stable in the IR.
While this leaves open the possibility that there are instabilities not localised in the IR (for example, instabilities of
the type studied by Gubser-Mitra \cite{Gubser:2000ec,Gubser:2000mm}), it is possible that for certain charges these are absent as well. These solutions would then 
provide the first top-down
examples of stable non-supersymmetric solutions with non-vanishing entropy in the IR. Note also that we can switch off one of the charges, leading to an $\eta$=1 geometry in the far IR in which there is also an emergent supersymmetry.

\subsection{Uplifted magnetic $\eta=1$ geometries}
We can uplift to $D=11$ the limiting $\eta=1$ geometry that appears at $T=0$ in the far IR. We again write
$Q=Q_{1}Q_{2}Q_{3}$ and find that as $r\to 0$
\begin{align}
&ds_{11}^{2}\approx \frac{\delta^{2/3}}{2^{1/12}Q^{1/4}}r^{1/3}\,\left[-U_{0}r\,dt^{2}+U_{0}^{-1}\frac{dr^{2}}{r^{2}}+\frac{1}{\left( 8Q\right)^{1/4}}\,\left(dx_{1}^{2}+dx_{2}^{2}\right) \right.\nn
&+\frac{2^{3/4}Q^{1/4}}{\delta}\,r^{-1}\,\left(d\mu_{4}^{2}+\mu_{4}^{2}\,d\phi_{4}^{2}\right)
+\left.\frac{2^{5/4}Q^{3/4}}{\delta}\,\sum_{i=1}^{3}Q^{-1}_{i}\,\left( d\mu_{i}^{2}+\mu_{i}^{2}\,\left(d\phi_{i}+2 A_{i} \right)^{2}\right)\right]
\end{align}
where we have defined $\delta=\sum_{i=1}^{3}Q_{i}\mu_{i}^{2}$ and the three non-vanishing magnetic gauge fields are given by
$A_{i}=-\varepsilon_{i}\frac{Q_{i}}{8\sqrt{Q}}\,\left(x_{1}\,dx_{2}-x_{2}\,dx_{1} \right)$ for $i=1,2,3$. In contrast to the uplifted
electric solutions given in \eqref{elul} we no longer see an $AdS_3$ factor.

\section{Supersymmetric magnetically charged black holes}\label{magsec2}
In this section we will discuss supersymmetric solutions of the $U(1)^4$ theory \eqref{eq:4dLag}
carrying three non-vanishing magnetic charges which approach 
$\eta$-geometries with $\eta=1$ in the far IR. Furthermore, these geometries
can be resolved, while preserving supersymmetry, by the addition of a fourth magnetic charge. If the fourth charge is small the solutions will have a large
an intermediate $\eta=1$ geometry scaling regime before approaching $AdS_2\times\mathbb{R}^2$ in the far IR.
Being supersymmetric these solutions are expected to be stable.

We consider magnetically charged solutions within the ansatz 
\begin{align}\label{eq:radans}
ds^{2}&=-e^{2W}\,dt^{2}+dr^{2}+e^{2U}\,\left(dx_{1}^{2}+dx_{2}^{2} \right)\,,\notag\\
F^{i}&=\tfrac{1}{2}\Bp q_{i}\,dx_{1}\wedge dx_{2}\,,\notag\\
\phi_{a}&=\phi_{a}(r)\,,
\end{align}
where $\Bp$ is a constant, given below, chosen to simplify some expressions.
In order to obtain supersymmetric solutions, following \cite{Duff:1999gh,Donos:2011pn} we will restrict our attention
to solutions with 
\begin{equation}
\sum_{i}q_{i}=0\,.
\end{equation}
It will be convenient to write
\begin{align}\label{eq:flux_constrains}
&q_{1}=Q+Z-\epsilon,\quad q_{2}=Q-Z-\epsilon\nn
&q_{3}=-2Q-\epsilon,\qquad q_{4}=3\,\epsilon\,.
\end{align}
As discussed in \cite{Donos:2011pn} the supersymmetry variations lead to the first order system of equations given by
\begin{align}\label{eq:SUSYflow}
-W^{\prime}+\frac{1}{2\sqrt{2}}\sum_{i}X_{i}+\frac{\alpha \Bp\Bp}{2\sqrt{2}}e^{-2U}\,\sum_{i}X_{i}^{-1}q_{i}=0\,,\notag\\
-U^{\prime}+\frac{1}{2\sqrt{2}}\sum_{i}X_{i}-\frac{\alpha \Bp}{2\sqrt{2}}e^{-2U}\,\sum_{i}X_{i}^{-1}q_{i}=0\,,\notag\\
-\sqrt{2}\phi_{a}^{\prime}-2\,\sum_{j}\,\partial_{\phi_{a}}X_{j}+2\alpha \Bp e^{-2U}\,\sum_{j}\,q_{j}\partial_{\phi_{a}}X_{j}^{-1}=0\,,
\end{align}
where $\alpha=\pm 1$.

We next recall that when all four charges are non-zero, there is a locus of supersymmetric magnetic $AdS_2\times\mathbb{R}^2$
solutions that we summarised in section \eqref{magads2sols}.

\subsection{Supersymmetric $\eta=1$ geometries in the IR}
Setting $\epsilon=0$ in \eqref{eq:flux_constrains} we have three non-vanishing charges and we can construct a
supersymmetric domain wall that approaches $AdS_4$ in the UV and an $\eta=1$ geometry in the IR. To see this
we can set up an approximate IR expansion to the equations \eqref{eq:SUSYflow}
of the form
\begin{align}
U&=\ln r+\dots\,,\qquad \quad W=3\,\ln r+\dots\,,\nn
\phi_{1}&=\ln\left(\frac{\left(3Q^{2}+Z^{2} \right)^{2}}{8\left(Q^{2}-Z^{2}\right)^{2}} \right)+2\,\ln r+\dots \,,\nn
\phi_{2}&=\ln\left(\frac{\left(3Q^{2}+Z^{2} \right)^{2}}{32Q^{2}\left(Q+Z\right)^{2}} \right)+2\,\ln r+\dots\,,\nn
\phi_{3}&=-\ln\left(\frac{\left(3Q^{2}+Z^{2} \right)^{2}}{32Q^{2}\left(Q-Z\right)^{2}} \right)-2\,\ln r+\dots\,,
\end{align}
where we have chosen the constant $\Bp$
\begin{align}
\Bp=\frac{16\,Q\,\left(Q-Z\right)\left(Q+Z\right)}{\left(3Q^{2}+Z^{2}\right)^{2}\alpha}\,.
\end{align}
This expansion yields the approximate metric behaviour
\begin{align}
ds_{4}^{2}\approx -r^{6}\,dt^{2}+dr^{2}+r^{2}\,\left(dx_{1}^{2}+dx_{2}^{2}\right)\,,
\end{align}
and after the coordinate transformation $r\rightarrow 2\rho^{-1/2}$ we obtain the metric \eqref{etageom} with $\eta=1$.

Using this expansion it is possible to construct supersymmetric solutions that approach $AdS_4$ in the UV and
this $\eta=1$ geometry in the IR. While this can be done directly, such solutions can also be obtained as a limit of the solutions that we construct in
the next subsection.

\subsection{Intermediate scaling in supersymmetric magnetic solutions}
We now construct supersymmetric magnetic solutions carrying four
magnetic charges which approach $AdS_4$ in the UV and $AdS_2\times\mathbb{R}^2$ in the IR with an
intermediate $\eta=1$ geometry scaling region.
We consider flows with the magnetic fluxes constrained as in \eqref{eq:flux_constrains} 
and we take $Z=0$ for simplicity. 
Recall from \eqref{firsttrunc} that these solutions lie within a consistent truncation with two scalars
$\phi_{1}$ and $\phi_{2}=-\phi_{3}$ and three gauge-fields. When $\epsilon\ne 0$ we have four non-vanishing magnetic charges and
we expect supersymmetric domain walls approaching $AdS_4$ in the UV and  $AdS_{2}\times\mathbb{R}^{2}$ in the IR. 
When $\epsilon$ is small there should be a large intermediate scaling $\eta=1$ geometry regime

The relevant supersymmetric $AdS_{2}\times\mathbb{R}^{2}$ solutions are given by
\begin{align}
W&=r/L,\qquad\qquad
L^2=\frac{2 e^{f_2}(1-e^{2 f_2})^2(1+e^{2 f_2})}{(3+2e^{2 f_2}+3 e^{4 f_2})^2}\nn
e^{U_{0}}&=\sqrt{6\Bp Q\alpha} \,\frac{e^{f_2/2} \sqrt{1-e^{2 f_2}} }{\sqrt{\left(3+6 e^{2 f_2}-e^{4 f_2}\right) g}}\,,\nn
e^{f_1}&=2\frac{\cosh(f_2)}{\sinh^2(f_2)},\qquad
\epsilon=\frac{\left(3+e^{2 f_2}\right) Q}{3+\cosh(2 f_2)-2 \sinh(2 f_2)}\,,
\end{align}
where $L$ is the radius of the $AdS_2$ and $f_1,f_2$ are the constant values of the scalars $\phi_1,\phi_2$, respectively.
This is a one-parameter family of solutions which we can take to be specified by $f_2$ (or by $\epsilon$).
Note that we will focus on $f_2<0$. These solutions have a ``universal" irrelevant operator with
dimension $\Delta=2$. There is also another irrelevant operator of dimension
\begin{equation}\label{eq:ir_dim}
\Delta_{IR}=\frac{6+2\,\cosh(2 f_{2})+\sqrt{-26+24\,\cosh(2 f_{2})+66\,\cosh(4 f_{2})}}{4+12\,\cosh(2 f_{2})}\,.
\end{equation}

We want to construct domain wall solutions that interpolate between these $AdS_2\times\mathbb{R}^2$ 
solution in the IR and $AdS_4$ in the UV.
In the UV we have the expansion
\begin{align}
W&=\frac{r}{L_{UV}}-\frac{1}{16}\,\left( 2\,c_{2}^{2}+c_{1}^{2}\right)\,e^{-2r/L_{UV}}+\ldots\nn
U&=\frac{r}{L_{UV}}-\frac{1}{16}\,\left( 2\,c_{2}^{2}+c_{1}^{2}\right)\,e^{-2r/L_{UV}}+\ldots\nn
f_{1}&=c_{1}\,e^{-r/L_{UV}}+\ldots\nn
f_{2}&=c_{2}\,e^{-r/L_{UV}}+\ldots
\end{align}
where $L_{UV}=1/\sqrt{2}$ and $c_{i}$ are two constants of integration.
For the IR expansion we have
\begin{align}
W&=W_{0}+r/L+c\,e^{r/L}+\ldots\nn
U&=U_{0}-\frac{9+20\,\cosh(2f_2)+3\,\cosh(4f_2)}{50+8\,\cosh(2f_2)+6\,\cosh(4\,f_2)}\,c\,e^{r/L}+\ldots\nn
\phi_1&=f_{1}+\frac{7+28\,\cosh(2\,f_2)-3\,\cosh(4f_2)}{25+4\,\cosh(2f_2)+3\,\cosh(4f_2)}\,c\,e^{r/L}+\ldots\nn
\phi_2&=f_2+8\sinh^{3}(f_2)\,\frac{1+3\cosh(2f_2)}{54\,\cosh(f_2)+7\,\cosh(3f_2)+3\,\cosh(5f_2)}\,c\,e^{r/L}+\ldots
\end{align}
where the constant $W_{0}$ corresponds to simple scalings of the time coordinate $t$ and $c$ is a deformation due to an irrelevant operator with $\Delta=2$. In this expansion we have not included the possibility for
a deformation of the operator with dimension given in \eqref{eq:ir_dim}. We could do this giving rise to additional
domain wall solutions.

\begin{figure}
\centering
\subfloat[Plot of $p_{1}$]{\includegraphics[height=4cm]{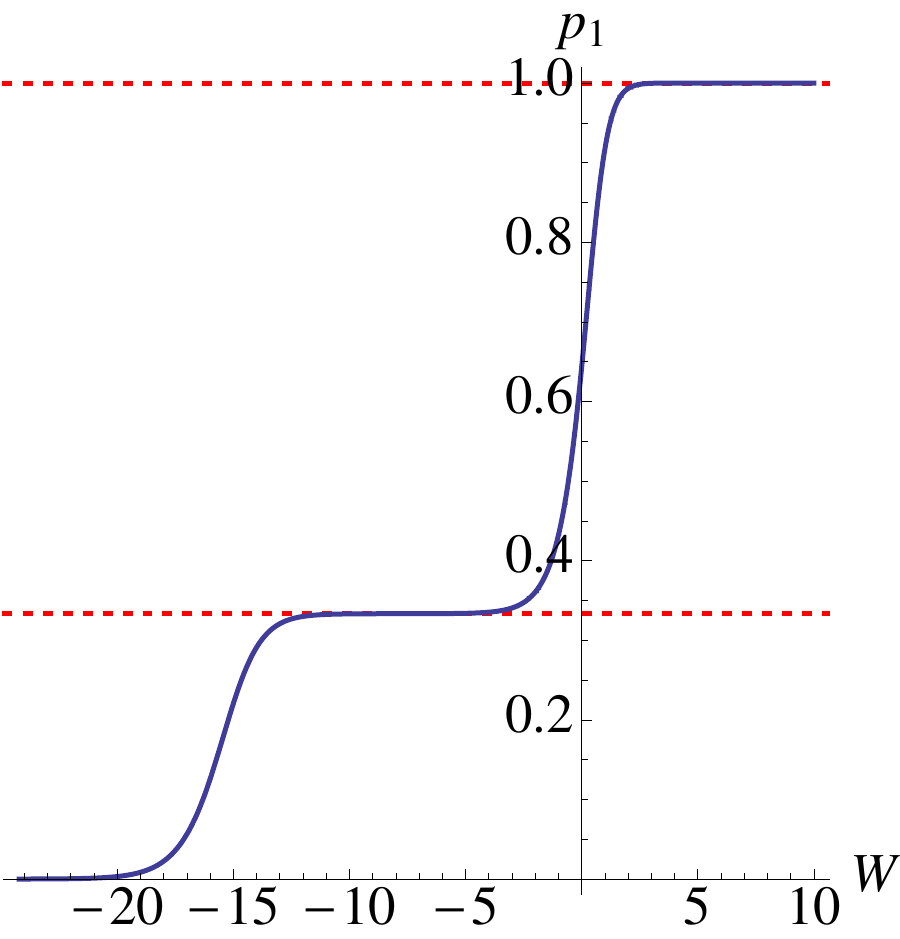}}\qquad\qquad
\subfloat[Plot of $p_{2}$]{\includegraphics[height=4cm]{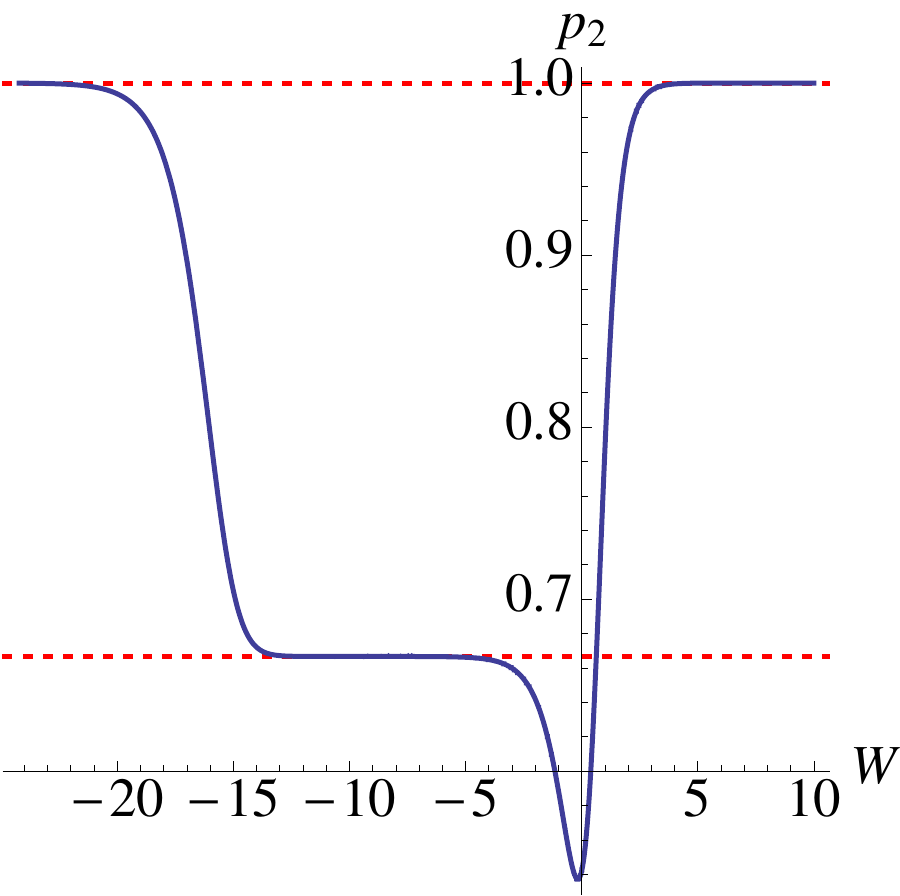}}\\
\subfloat[Plot of $p_{3}$]{\includegraphics[height=4cm]{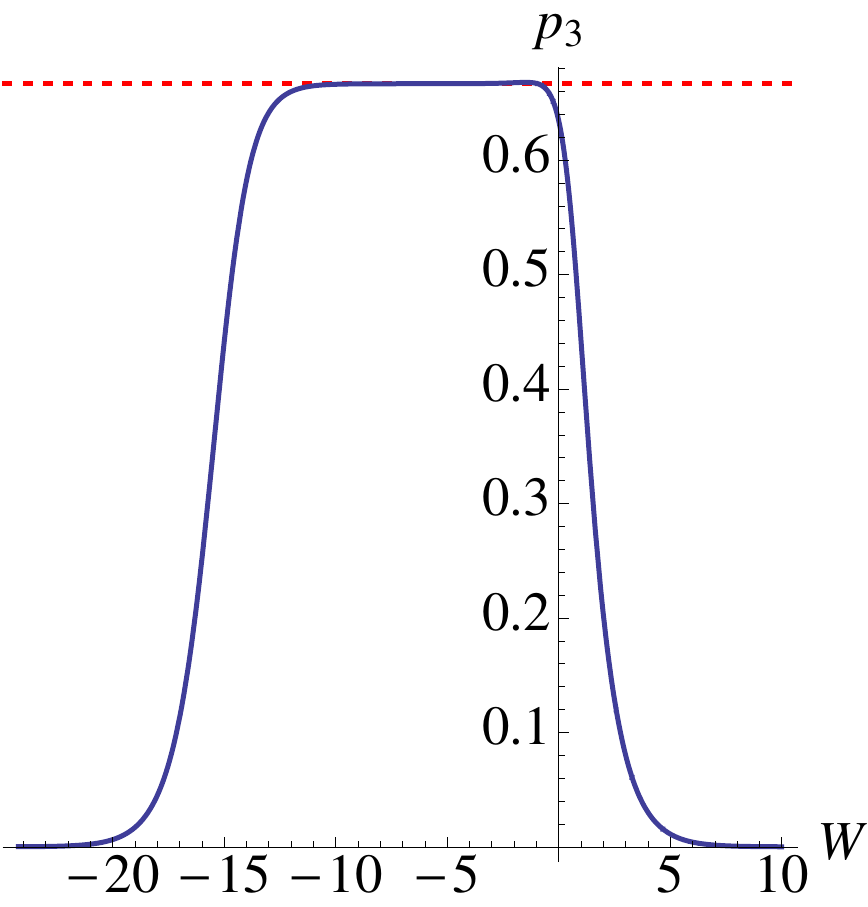}}\qquad\qquad
\subfloat[Plot of $p_{4}$]{\includegraphics[height=4cm]{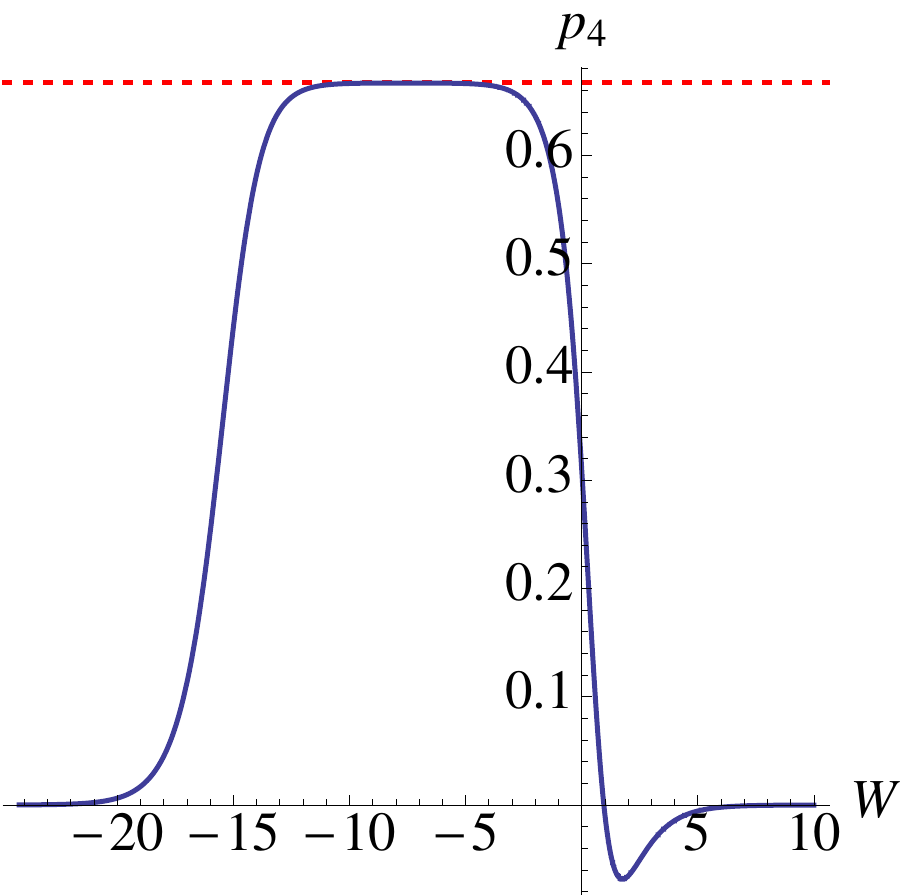}}
\caption{A plot of the four functions $p_i$, defined in \eqref{defpees}, as a function of
the radial coordinate $W$ for supersymmetric magnetically charged solutions. 
The plots reveal three scaling regimes, corresponding to $AdS_4$ for
large $W$, an $\eta=1$ geometry for intermediate $W$ and $AdS_2\times\mathbb{R}^2$ for
small $W$.}\label{fig:susy_plots}
\end{figure}

We choose $\epsilon/Q=2\times 10^{-10}$. We have two integration
constants in the IR and two integration constants in the UV. Since we have set 
$\phi_{2}=-\phi_{3}$ we have four first order BPS equations to solve, given in
\eqref{eq:SUSYflow}, and so we expect to find a unique solution.
Indeed we constructed such a solution numerically. To discuss the scaling properties of the solution
it is convenient to define
\begin{align}\label{defpees}
p_{1}&=\frac{U^{\prime}}{W^{\prime}},\qquad p_{2}=1+\frac{W^{\prime\prime}}{W^{\prime\,2}}\,,\nn
p_{3}&=\frac{\phi_{2}^{\prime}}{W^{\prime}},\qquad p_{4}=\frac{\phi_{1}^{\prime}}{W^{\prime}}\,,
\end{align}
and consider the $p_i$ to be functions of $W$, which is natural if we decided to use $W$ as a radial coordinate instead of
$\rho$ in our ansatz \eqref{eq:radans}.
Corresponding to the three different scaling regimes we expect to see
\begin{itemize}
\item $AdS_{2}\times\mathbb{R}^{2}$
\begin{equation}
p_{1}=0,\quad p_{2}=1\quad p_{3}=p_{4}=0\nn
\end{equation}
\item $\eta=1$ geometry
\begin{equation}
p_{1}=1/3,\quad p_{2}=2/3,\quad p_{3}=p_{4}=2/3\nn
\end{equation}
\item $AdS_{4}$
\begin{equation}
p_{1}=1,\quad p_{2}=1,\quad p_{3}=p_{4}=0\,.
\end{equation}
\end{itemize}
In figure \ref{fig:susy_plots} we have plotted the functions $p_i(W)$, which clearly reveals these three regimes.

We conclude this section by noting that all of the supersymmetric solutions with magnetic charges that we have
constructed in this section have electrically charged analogues obtained from the duality transformation
\eqref{dual}. However, these electrically charged solutions will not be supersymmetric.

\section{$\eta$-geometries from dimensional reduction}\label{dimredsec}
We finish this paper by describing some simple ways in which the 
$\eta$-geometries \eqref{etageom} can be obtained via Klauza-Klein reduction.
Similar observations were made earlier (independently) in the context of specific classes of models in
\cite{Gouteraux:2011ce}. For example, to obtain $\eta$ geometries in $D=4$
we start with $AdS_{2+k}\times\mathbb{R}^2$, given in Poincar\'e type coordinates by
\begin{align}
ds^2&=L^2\left[-\frac{dt^2}{r^2}+\frac{dr^2}{r^2}+\frac{dy_a dy_a}{r^2}\right]+ dx_1^2+dx_2^2\,,
\end{align}
where $L$ is the radius of the $AdS_{2+k}$ and $a=1,\dots ,k$.
We now perform a dimensional reduction on the $k$ spatial dimensions $y_a$. To do this we rewrite the metric in the form
\begin{align}
ds^2&=r^k\left(\frac{1}{r^k}\left\{L^2\left[-\frac{dt^2}{r^2}+\frac{dr^2}{r^2}\right]+ dx_1^2+dx_2^2\right\}\right)+\frac{L^2}{r^2}dy_a dy_a\,.
\end{align}
A straightforward calculation shows that the metric in the round braces is the $D=4$ Einstein-frame metric, and we see
an $\eta$-geometry with $\eta=k$ and $\ell=L$. 
A simple extension is to replace $AdS_{2+k}$ with a Lifshitz geometry with dynamical exponent $z$. We then have
\begin{align}
ds^2&=r^k\left(\frac{1}{r^k}\left\{L^2\left[-\frac{dt^2}{r^{2z}}+\frac{dr^2}{r^2}\right]+ dx_1^2+dx_2^2\right\}\right)+\frac{L^2}{r^2}dy_a dy_a\,.   
\end{align}
We again reduce on the $k$ spatial dimensions $y_a$ and perform a coordinate transformation to find an
$\eta$ geometry in  $D=4$ with $\eta=k/z$ and $\ell=L/z$.

These constructions immediately provide rich top-down constructions. For example, we can start with the $AdS_3\times \mathbb{R}^2$
solutions of $D=5$ maximal gauged supergravity that are supported by magnetic fields and, in general, scalar fields which were studied in 
\cite{Almuhairi:2011ws,Donos:2011pn}. 
These can be uplifted on an $S^5$ to obtain exact solutions
of type IIB supergravity. A subclass of solutions are also solutions of Romans $D=5$ gauged supergravity and furthermore there is
a unique solution which is a solution of $D=5$ minimal gauged supergravity, and these can be uplifted
to both type IIB and $D=11$ in infinite numbers of ways \cite{Gauntlett:2006ai,Gauntlett:2007ma,Gauntlett:2007sm} . After dimensional reduction on a spatial dimension contained within
the $AdS_3$ factor we obtain infinite top-down examples of $\eta$ geometries in $D=4$ with $\eta=1$ that are supported by magnetic fields
as well as other scalar fields.
Interestingly, for the solutions in maximal gauged supergravity and in Romans
supergravity there is a supersymmetric locus of solutions and this provides supersymmetric examples of $\eta=1$ geometries in $D=4$.

These constructions might provide a helpful framework for obtaining useful insights into the holographic dictionary for $\eta$-geometries 
along the lines of \cite{Gouteraux:2011qh}.

\subsection*{Acknowledgements}
We thank S. Hartnoll for helpful discussions.
CP is supported by an I.K.Y. Scholarship.
This work was supported in part by STFC grant ST/J0003533/1.

\appendix

\providecommand{\href}[2]{#2}\begingroup\raggedright\endgroup

\end{document}